\documentclass[aps,prb,showpacs,superscriptaddress,floatfix]{revtex4}
\usepackage{amsmath}
\usepackage{citesort}
\usepackage{graphicx}
\begin{document}
\title{
  Residual absorption at zero temperature in $d$-wave superconductors
  }
\author{E. Schachinger}
\email{schachinger@itp.tu-graz.ac.at}
\homepage{www.itp.tu-graz.ac.at/~ewald}
\affiliation{Institut f\"ur Theoretische Physik, Technische Universit\"at
Graz\\A-8010 Graz, Austria}
\author{J.P. Carbotte}
\affiliation{Department of Physics and Astronomy, McMaster University,\\
Hamilton, Ontario, L8S 4M1 Canada}
\date{\today}
\begin{abstract}
  In a $d$-wave superconductor with elastic impurity scattering,
  not all the available optical spectral weight goes into the
  condensate at zero temperature, and this leads to residual absorption.
  We find that for a range of impurity parameters
  in the intermediate coupling regime between Born (weak) and
  unitary (strong) limit, significant oscillator strength remains
  which exhibits a cusp like behavior of the real part of the
  optical conductivity with upward curvature
  as a function of frequency, as well as a quasilinear
  temperature dependence of the superfluid density. The
  calculations offer an explanation of recent data on 
  ortho-II YBa$_2$Cu$_3$O$_{6.5}$ which has been considered
  anomalous.
\end{abstract}
\maketitle
\newpage
\section{Introduction}

Recent measurements\cite{turner} of the microwave conductivity
up to $21\,$GHz at low temperatures in single crystal oxygen
ordered ortho-II YBa$_2$Cu$_3$O$_{6.5}$ (YBCO) have revealed that there
remains significant absorption at $T=0$ in the superconducting
state, while at the same time the inverse square of the penetration
depth shows a linear in temperature behavior.
In the regime where this happens, the real part of the optical
conductivity $\sigma_1(\omega)$ shows an unusual characteristic
upward curvature very different from a Drude-like form.
Anomalous residual absorption has also been seen in the past
in THz work on thin films of
Bi$_2$Sr$_2$CaCu$_2$O$_{8+\delta}$ (Bi2212)\cite{corson} and in this
case was interpreted as due to coupling to a collective
mode, with the spectral weight of this mode tracking the
temperature dependence of the superfluid density.

In a pure $d$-wave BCS superconductor all of the optical spectral
weight will go into the condensate at $T=0$ providing a delta
function to the real part of the conductivity at zero frequency.
This follows from the Ferrell-Glover-Tinkham sum rule\cite{ferrell,%
tinkham} which holds when the kinetic energy change between
normal and superconducting state is negligible. When elastic
impurity scattering is introduced into the system, this no
longer holds. Some optical spectral weight will remain
at finite frequency under
the real part of the conductivity at $T=0$ and correspondingly,
the superfluid density will be reduced below its pure limit
value. At the same time, however, in the unitary (strong) limit
of impurity scattering which is the most studied case so far, a
switch from linear to quadratic behavior in temperature is
expected for the penetration depth. In this work we consider
the case when the impurity scattering strength falls in the
intermediate regime between Born (weak) and unitary (strong)
limit. We find that for a range of impurity parameters,
i.e.: concentration and strength of the potential measured in
terms of the width of the electronic band, the residual, impurity induced
absorption remains very significant, while at the same time
the superfluid density obeys a quasilinear law over the entire
low temperature range and, more importantly, $\sigma_1(\omega)$
shows anomalous behavior in agreement with the experimental
observation in ortho-II YBCO.

\section{Formalism}

The general expression for the optical conductivity at
temperature $T$ and frequency $\nu$ is well known and has been
given in many places.\cite{mars1,schur,schach1,schach2,hirschf1}
It can be written as:
\begin{widetext}
\begin{equation}
  \sigma(T,\nu) = -\frac{\Omega^2_p}{4\pi}\left\langle\left[
      -\int\limits_0^\infty\!d\omega\,\tanh\left(\frac{\beta\omega}{2}
        \right)J_\theta(\omega,\nu)
     +\int\limits_{-\nu}^\infty\!d\omega\,
        \tanh\left(\frac{\beta(\omega+\nu)}{2}\right)
        J_\theta(-\omega-\nu,\nu)
      \right]\right\rangle_\theta,
  \label{eq:1}
\end{equation}
where $\Omega_p$ is the plasma frequency and the function
$J_\theta(\omega,\nu)$ takes on the form
\begin{eqnarray}
  2J_\theta(\omega,\nu) &=& \frac{1}{E_1+E_2}\left[1-
    N(\theta,\omega)N(\theta,\omega+\nu)
    -P(\theta,\omega)P(\theta,\omega+\nu)\right]\nonumber\\
  &&+\frac{1}{E^\ast_1-E_2}\left[1+
     N^\ast(\theta,\omega)N(\theta,\omega+\nu)
    +P^\ast(\theta,\omega)P(\theta,\omega+\nu)\right].
  \label{eq:2}
\end{eqnarray}
\end{widetext}
In Eq.~(\ref{eq:1}) $\beta = 1/k_B T$, with $k_B$ the Boltzmann factor
and $T$ the temperature. In Eq.~(\ref{eq:2})
\begin{eqnarray*}
  E_1(\theta,\omega) &=& \sqrt{\tilde{\omega}^2(\omega+i0^+)-
                 \tilde{\Delta}^2(\theta,\omega)},\\
  E_2(\theta,\omega,\nu) &=& E_1(\theta,\omega+\nu),
\end{eqnarray*}
and
\[
  N(\theta,\omega) = \frac{\tilde{\omega}(\omega+i0^+)}
          {E_1(\theta,\omega)},
  \qquad P(\theta,\omega) = \frac{\tilde{\Delta}(\theta,\omega)}
          {E_1(\theta,\omega)}.
\]
Here $\theta$ is the angle
giving the direction of momentum on the Fermi surface in the
two dimensional CuO$_2$ Brillouin zone. In BCS the gap does not
depend on frequency $\omega$ and has the form
$\tilde{\Delta}(\theta,\omega) = \Delta(T)\cos(2\theta)$ with
$\Delta(T)$ a temperature dependent amplitude. The angular
bracket $\langle\cdots\rangle_\theta$ indicates an average over
angle $\theta$.

The renormalized frequencies $\tilde{\omega}(\omega)$ do not
depend on angle but are
changed because of the elastic impurity scattering which is the
only scattering processes accounted for. The
equation for $\tilde{\omega}(\omega+i0^+)$ in a $t$-matrix
approximation is\cite{schur,schach1,schach2,hirschf1,hirschf2}
\begin{equation}
  \label{eq:3}
  \tilde{\omega}(\omega+i0^+) =
  \omega+i\pi\Gamma^+\frac{\left\langle 
  N(\theta,\omega)\right\rangle_\theta}
    {c^2+\left\langle N(\theta,\omega)\right\rangle^2_\theta}.
\end{equation}
$\Gamma^+$ is
proportional to the impurity concentration and
$c=1/\left(2\pi N(0)V_{imp}\right)$
with $N(0)$ the normal band density of states at the
Fermi surface and $V_{imp}$ the impurity potential.
An important parameter is $\gamma$ defined as
$\Im{\rm m}\tilde{\omega}(\omega+i0^+)$ at $\omega=0$ which is
shown in the bottom frame of Fig.~\ref{fig:4}.
For small values of $V_{imp}$, Born's approximation applies and
$c\to\infty$ with $t^+ = \Gamma^+/c^2$ and the impurity term
in Eq.~(\ref{eq:3}) becomes proportional to
$\langle N(\theta,\omega)\rangle_\theta$, its real part is
the superconducting state density of states
$(N_s(\omega))$, while for large
values of $V_{imp}$, $(c\to 0)$ it is
inversely proportional to $\left\langle%
N(\theta,\omega)\right\rangle_\theta)$
(unitary limit). Both Born and unitary limit are extreme and a
finite $c$ is more realistic.
As $N(0)$ is roughly inversely proportional to the
electronic band width $(W)$, we have $V_{imp} \simeq W/(2\pi c)$.
For $c=0.2$ the
corresponding $V_{imp}$ is 1.3 times the band width.

\section{Results and Discussion}

The quasiparticle density of states $N_s(\omega)$ for a $d$-wave
superconductor in the case of a finite value of $c=0.2$ is shown
\begin{figure}
  \vspace*{-5mm}
  \includegraphics[width=8cm]{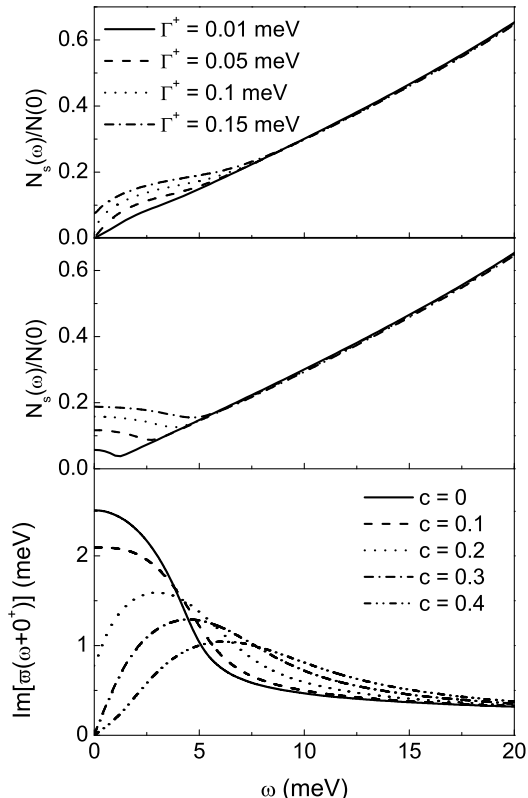}
  \vspace*{-8mm}
  \caption{The quasiparticle density of states $N_s(\omega)/N(0)$
    vs $\omega$, $\Delta_0 = 24\sqrt{2}\,$meV. The top
    frame is for $c=0.2$, and the middle frame
    is for the unitary limit, $c=0$. The bottom frame gives
    $\Im{\rm m}[\tilde{\omega}(\omega+0^+)]$ vs $\omega$ for
    $\Gamma^+ = 0.15\,$meV and various values of $c$.
    }
  \label{fig:1}
\end{figure}
in Fig.~\ref{fig:1} (top frame).
The gap amplitude was set at $24\sqrt{2}\,$meV
and the impurity parameter $\Gamma^+$ was varied from $\Gamma^+%
= 0.01\,$meV (solid), $\Gamma^+=0.05\,$meV (dashed),
$\Gamma^+=0.1\,$meV (dotted), and $\Gamma^+=0.15\,$meV (dash-dotted).
Clearly, impurities affect most strongly
the low energy region near $\omega = 0$ up to, say, $10\,$meV.
The pure
limit, a straight line through the origin is approached as
$\Gamma^+$ is reduced, but even for the smallest
value of $\Gamma^+ = 0.01\,$meV,
changes in $N_s(\omega)$ persist up to $\omega\simeq 4\,$meV.
It is the behavior of $N_s(\omega)$ in the low
energy range which determines
low temperature properties such as the specific heat.
The behavior found for a finite $c$
could not be inferred from a knowledge of the
unitary or Born limit. In the Born limit the deviation from
linearity would only occur at an exponentially small value
of $\omega$. Results for the unitary limit are shown in the
middle frame and are seen to be very different from the top frame.
The energy dependence of the
quasiparticle lifetime of Eq.~(\ref{eq:3}) $\Im{\rm m}\,%
\tilde{\omega}(\omega+i0^+)$ is shown in the bottom frame of
Fig.~\ref{fig:1}. For $c=0$, the scattering rate has its maximum
at $\omega=0$ and then drops monotonically as $\omega$ increases.
For the larger values of $c$ shown here this is no longer the
case. A maximum in scattering rate occurs around
$\omega\stackrel{<}{\sim}8\,$meV which is much larger than
its value at $\omega=0$ (denoted $\gamma$).

\begin{figure}
  \vspace*{-5mm}
  \includegraphics[width=8cm]{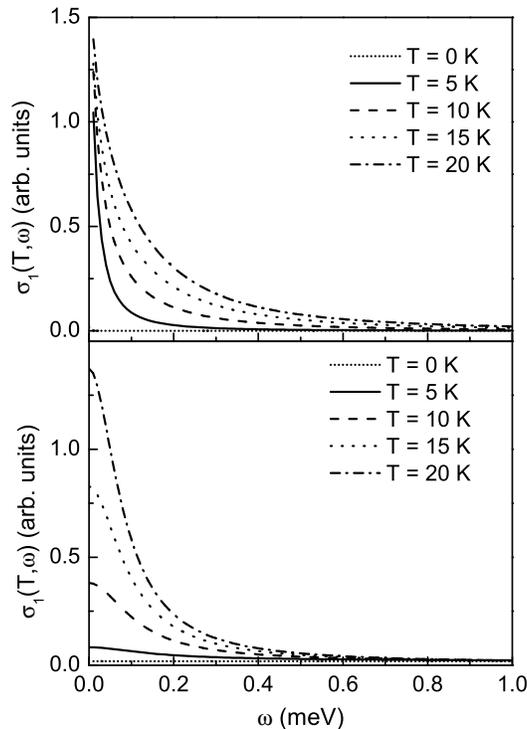}
  \vspace*{-8mm}
  \caption{The real part of the optical conductivity
    $\sigma_1(T,\omega)$ vs $\omega$
    for five temperatures. The gap amplitude
    $\Delta_0 = 24\sqrt{2}\,$meV and the impurity parameter
    $\Gamma^+=0.01\,$meV. The top frame is for $c=0.3$ and the
    bottom for $c=0$.
    The computer units on the conductivity are such that
    a factor of $\Omega_p^2/8\pi$ needs to be applied.
    }
  \label{fig:2}
\end{figure}

\begin{figure}
  \vspace*{-5mm}
  \includegraphics[width=8cm]{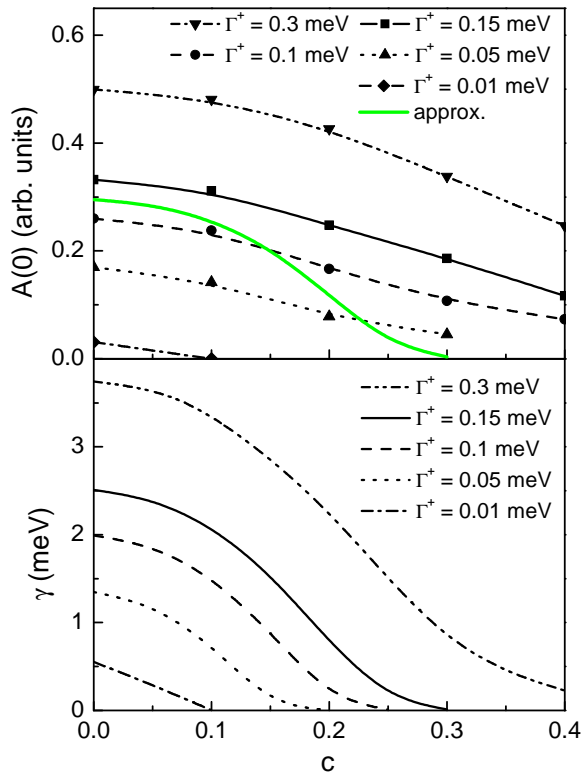}
  \vspace*{-8mm}
  \caption{Top frame: remaining
    spectral weight $A(0)=\int_{0+}^\infty d\omega\,\sigma_1(0,\omega)$
    as a function of the impurity potential strength $c$.
    The bottom frame gives the zero
    frequency value of the effective scattering in the superconducting
    state $\gamma(c)$ as a function of $c$. The solid gray line
    represents the results of the approximate formula for
    $\Gamma^+ = 0.15\,$meV.
    }
  \label{fig:4}
\end{figure}

The real part of the optical conductivity for a $d$-wave BCS
superconductor is shown in Fig.~\ref{fig:2}
in units of $\Omega^2_p/8\pi$.
Five values of temperature are considered, namely $T=0\,$K
(short dotted line) $T=5\,$K (solid),
$T=10\,$K (dashed), $T=15\,$K (dotted), and
$T=20\,$K (dash-dotted). The $d$-wave gap amplitude was set
at $24\sqrt{2}\,$meV. The top frame applies to an intermediate
scattering case with $\Gamma^+ = 0.01\,$meV and $c=0.3$. The
bottom frame is for comparison and shows unitary limit results
$(c=0)$. The real part of the optical conductivity
$\sigma_1(T,\omega)$ vs $\omega$ up
to $1\,$meV, is radically different in the two cases. For
$c=0.3$ the finite $T$ curves are concave up and
drop rapidly with increasing $\omega$
and show a very narrow half-width in qualitative
agreement with the experiments
by Turner {\it et al.}\cite{turner} and Hosseini {\it et al.}%
\cite{hoss} The concave upward nature of these curves is due
to the energy dependence of the quasiparticle
scattering rate obtained from Eq.~(\ref{eq:3}).
These curves all correspond to the temperature dominated
regime $T\gg\gamma$ (see Fig.~\ref{fig:4}, bottom frame). Their
half width is not related to the value of $\gamma$ but rather to
temperature which samples the important energy dependence
in $\Im{\rm m}\,\tilde{\omega}(\omega+i0^+)$. The
$T=0$ curve is impurity dominated and,
consequently, shows a qualitatively different behavior.
It cannot be obtained from an extrapolation to $T=0$ of the
finite $T$ curves shown.
These conductivity curves are in qualitative
agreement with the experimental findings.\cite{turner}
In the unitary limit the curves show instead a concave
downward behavior at small $\omega$ and a broad half-width.
This can be traced to
the fact that the zero frequency scattering rate
$\gamma$, is a rapidly increasing
function of decreasing $c$ as is seen in the bottom
frame of Fig.~\ref{fig:4}. For the unitary limit $\gamma$ is
greater than its normal state value
$\pi\Gamma^+$ and is given
by the well known relation $\gamma\simeq 0.63\sqrt{\pi\Delta_0%
 \Gamma^+}$ while in the Born limit $\gamma$ is exponentially
dependent on $\Gamma^+$ ($\gamma(c\to \infty) = 4\Delta_0\exp(-%
\Delta_0/2\Gamma_N)$ with $\Gamma_N = \Gamma^+/c^2$.)

The quantity of interest is the total spectral weight
remaining under
\begin{figure}
  \vspace*{-5mm}
  \includegraphics[width=8cm]{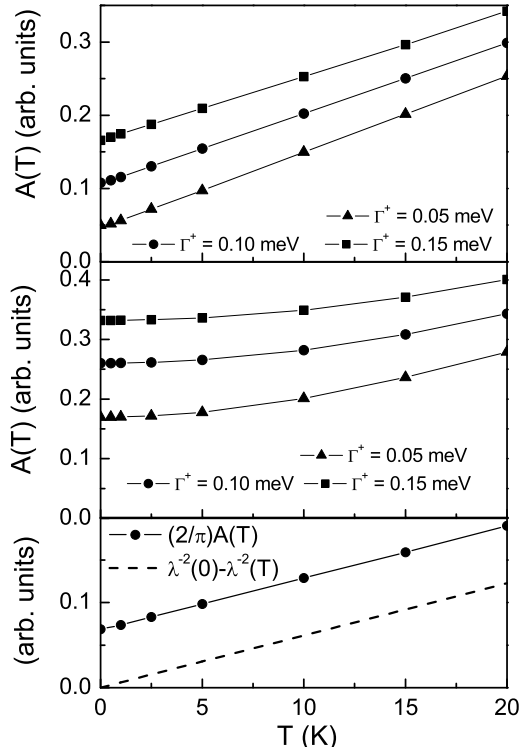}
  \vspace*{-8mm}
  \caption{Top frame: the remaining spectral weight
    $A(T) = \int_{0^+}^\infty d\omega\,\sigma_1(T,\omega)$ as a
    function of temperature $T$ in the range 0 to $20\,$K for
    gap $\Delta_0 =%
    24\sqrt{2}\,$meV. The impurity potential is characterized
    by $c=0.3$ (to be compared with $c=0$ for the unitary
    limit in the middle frame).
    The bottom
    frame compares $(2/\pi)A(T)$ with $\lambda^{-2}(0)-%
    \lambda^{-2}(T)$ for the case $\Gamma^+ = 0.1\,$meV and
    $c=0.3$.
    }
  \label{fig:3}
\end{figure}
the real part of the optical conductivity at temperature $T$,
  $A(T) = \int_{0^+}^\infty\!d\omega\,\sigma_1(T,\omega)$.
Numerical results are presented in Fig.~\ref{fig:3} which
consists of three frames. The top frame is for
$c=0.3$ while the middle frame, included for comparison,
is for the unitary limit. In both frames $\Delta_0=%
24\sqrt{2}\,$meV and the temperature range considered is up to
$20\,$K $(t=T/T_c\simeq 0.11)$.
The solid squares are
for $\Gamma^+ = 0.15\,$meV, solid circles $\Gamma^+ = 0.1\,$meV,
and solid triangles $\Gamma^+ = 0.05\,$meV.
The curves
show a quasi linear temperature dependence over the entire
temperature range considered.
This is in sharp contrast to the
results in the middle frame. The temperature dependence is now
quadratic which is the behavior expected in the unitary
limit. In the bottom frame we
compare $A(T)$ vs $T$ with the corresponding results for the
penetration depth $\lambda^{-2}(0)-\lambda^{-2}(T)$ vs $T$. We
see that the two curves are parallel and that the dotted curve
goes through zero at the origin. This follows directly from
the Ferrell-Glover-Tinkham sum rule which
is satisfied to within our numerical accuracy. In our computer
units the sum rule is
\begin{equation}
  \label{eq:6}
  \lim_{\omega\to 0}\omega\sigma_2(T,\omega) +
  \frac{2}{\pi}\int\limits_{0^+}^\infty\!d\omega\,
  \sigma_1(T,\omega) = 2,
\end{equation}
where $\sigma_2$ is the imaginary part of the conductivity.
While we have not attempted a best fit to experiment,
the results of the bottom frame of
Fig.~\ref{fig:3} are in qualitative agreement with the findings
of Turner {\it et al.}\cite{turner}

Zero temperature values for
$A(0)$ are shown in the top frame of Fig.~\ref{fig:4}. The lines
are a guide to the eye through numerical results
based on full conductivity calculations. The
points are solid down-triangles for $\Gamma^+ = 0.3\,$meV,
 solid squares for $\Gamma^+ = 0.15\,$meV, solid
circles for $\Gamma^+ = 0.1\,$meV, solid up-triangles for
$\Gamma^+ = 0.05\,$meV, and solid diamonds for
$\Gamma^+ = 0.01\,$meV.
In all cases $A(0)$, which is entirely due to impurities,
drops with increasing $c$. It is
largest in the unitary limit but remains significant
even when $c$ has increased to 0.4 provided $\Gamma^+$ is not too small.
In the bottom frame
we show $\gamma(c)$ vs $c$, and stress that it becomes small
more rapidly than does $A(0)$ as $c$ increases; consequently,  significant
absorption can remain even with small $\gamma$. As can be seen
in the top frame of Fig.~\ref{fig:2}, the value of $A(0)$,
which gives the optical weight under
$\sigma_1(\omega)$,
samples importantly the quasiparticle scattering rate away
from $\omega=0$ and, thus, $A(0)$ is not directly related to
$\gamma(c)$. There is much more absorption than would be
indicated by the value of $\gamma(c)$ at the larger values of
$c$ considered here. This is emphasized by the solid gray line
in the top frame of Fig.~\ref{fig:4} which corresponds to the
results of the approximate formula
$A(0) \simeq (\gamma/\Delta_0)\ln(4\Delta_0/\gamma)$
obtained by approximating the $\omega$-dependent
quasiparticle scattering rate in the exact expression by its
$\omega=0$ value, i.e.: by $\gamma(c)$. Here $\Gamma^+ =%
0.15\,$meV. Comparison with the full results (solid squares)
reveals that the agreement is reasonable for $c\le 0.1$, i.e.:
the region for which the quasiparticle scattering rate has its
maximum at $\omega = 0$, but fails when its maximum falls
instead at finite $\omega$ (see Fig.~\ref{fig:1}, bottom frame).

In their analysis of optical conductivity of thin films of
Bi2212 Corson {\it et al.}\cite{corson}
found a remaining spectral weight at $T=0$ of about 30\% of the
corresponding value of the superfluid density. In our language
this corresponds to a value of $A(0)\stackrel{>}{\sim} 0.5$.
Corson {\it et al.}\cite{corson} also found that the remaining
optical weight was distributed over a width of order $12\,$K
independent of temperature. Reference to the theoretical curves
in Fig.~\ref{fig:4} shows that this value of $A(0)$ (top frame)
is rather high for the observed value of the width $\sim\gamma$
(bottom frame); the data suggests $c\stackrel{>}{\sim} 0.4$
and $\Gamma^+$ of the same order as was
required\cite{schach3} to fit
the single crystal infrared optical data of Tu {\it et al.}\cite{tu}
in Bi2212, $\Gamma^+ = 0.61\,$meV. While
Corson {\it et al.} interpret their data in terms of extra
absorption due to a collective mode, part if not all is due
instead to the impurity induced absorption
discussed in this paper.

We mention previous work on film of YBa$_2$Cu$_3$O$_{7-\delta}$
by Hensen {\it et al.}\cite{hensen} in which the sizable surface
resistance observed at low temperatures and attendant quasi
linear temperature dependent superfluid density is seen to
be consistent with $d$-wave and impurity scattering with a
phase shift intermediate between Born and unitary limit. This
work, however, is limited to the consideration of a single
microwave frequency, namely $87\,$GHz, and thus remains silent
on the unusual frequency dependence of $\sigma_1(\omega)$ vs
$\omega$ found also to be the characteristic of this regime
in our theoretical work and in the experimental data of
Turner {\it et al.}\cite{turner}

\section{Conclusion}

For impurity scattering with a phase shift intermediate between
0 and $\pi/2$ we have found a regime in which considerable
residual absorption can remain at zero temperature while at
the same time the superfluid density is quasilinear in
temperature. In addition, and more importantly, in this
regime, the real part of the optical conductivity, $\sigma_1%
(\omega)$, exhibits a distinctly non-Drude dependence on
frequency in the microwave region. At low, but finite, temperatures
$\sigma_1(\omega)$ vs $\omega$ has a cusp like dependence on
$\omega$ out of $\omega=0$, followed by a rapid drop
(concave upward curvature) as $\omega$ increases. The width
at half maximum of $\sigma_1(\omega)$, which can be very small,
varies strongly with temperature. This behavior is traced to
the strong energy dependence of quasiparticle lifetime which
is sampled in the temperature dominated regime discussed in
this work. All aspects of this work are in good qualitative
agreement with recent data in ortho-II YBa$_2$Cu$_3$O$_{6.5}$.

\begin{acknowledgments}
Research supported by the Natural Sciences and Engineering
Research Council of Canada (NSERC) and by the Canadian
Institute for Advanced Research (CIAR). J.P.C. thanks D.M.~Broun
for discussions.
\end{acknowledgments}

\end{document}